\documentclass{article} 

\usepackage{graphicx}
\usepackage{amsmath}
\usepackage{amssymb}
\usepackage{color}
\usepackage[numbers]{natbib}
\usepackage{hyperref}
\usepackage{caption}

\hypersetup{colorlinks=true,linkcolor=blue}

\title{Efficient cruising for swimming and flying animals is dictated by fluid drag}
\author{
  Daniel Floryan%
    \thanks{Mechanical and Aerospace Engineering, Princeton University, Princeton NJ 08544, USA}
    \thanks{Email address for correspondence: \href{mailto:dfloryan@princeton.edu}{dfloryan@princeton.edu} }
  \and Tyler Van Buren%
  \footnotemark[1]\
  \and Alexander J. Smits%
  \footnotemark[1]\
 }

\begin{document}
\date{}
\maketitle

\begin{abstract}

Many swimming and flying animals are observed to cruise in a narrow range of Strouhal numbers, where the Strouhal number ${St = 2fA/U}$ is a dimensionless parameter that relates stroke frequency $f$, amplitude $A$, and forward speed $U$. Dolphins, sharks, bony fish, birds, bats, and insects typically cruise in the range $0.2 < St < 0.4$, which coincides with the Strouhal number range for maximum efficiency as found by experiments on heaving and pitching airfoils. It has therefore been postulated that natural selection has tuned animals to use this range of Strouhal numbers because it confers high efficiency, but the reason why this is so is still unclear. Here, by using simple scaling arguments, we argue that the Strouhal number for peak efficiency is largely determined by fluid drag on the fins and wings. 

\end{abstract}

Swimming and flying animals across many species and scales cruise in a relatively narrow range of Strouhal numbers $0.2 < St < 0.4$  \cite{triantafyllou1991wake, taylor2003flying}. The Strouhal number $St = 2f A/U$ is a dimensionless parameter that relates stroke frequency $f$, stroke amplitude $A$, and forward speed $U$. It has been hypothesized that for animals that range widely or migrate over long distances, natural selection should favor swimming and flying motions of high propulsive efficiency, and so the kinematics, described by the Strouhal number, should be tuned for high propulsive efficiency. Indeed, the cruising range of Strouhal numbers observed in nature overlaps the range of Strouhal numbers experimentally shown to result in high propulsive efficiency for simple propulsors \cite{triantafyllou1991wake, floryan2017scaling, van2018scaling}.  

\begin{figure}
  \centering
  \includegraphics[width=0.5\linewidth]{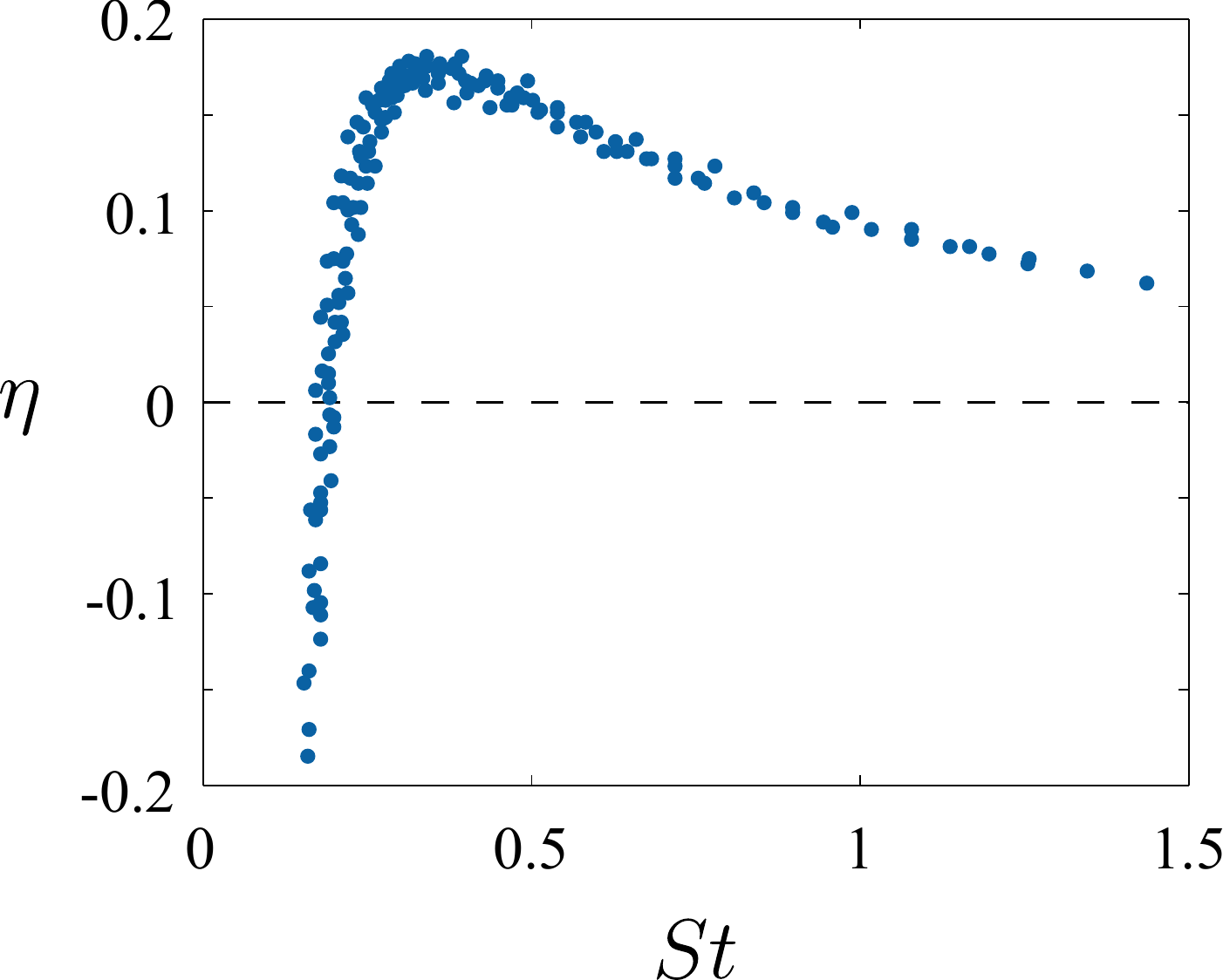}
  \caption{A typical efficiency curve showing efficiency $\eta$ as a function of $St$. Data are for a heaving and pitching NACA0012 foil \cite{quinn2015maximizing} ($A/L= 0.19$, and heaving leads pitching by $90^\circ$).}
  \label{fig:eff}
\end{figure}

A typical efficiency curve for a simple propulsor is shown in figure~\ref{fig:eff}. 
We see that at low Strouhal numbers, the efficiency rapidly rises with increasing Strouhal number, reaches a maximum, and then falls off relatively slowly with further increases in Strouhal number.  Here, the propulsive efficiency $\eta$ is defined as $\eta = TU/P$, where $T$ is the mean net thrust that propels the animal forward, $U$ is the mean forward cruising speed, and $P$ is the mean mechanical power required to create the thrust. 

What dictates the Strouhal number that leads to maximum efficiency? Three prevailing theories have been proposed. The first  \cite{triantafyllou1991wake, triantafyllou1993optimal} argues that peak efficiency occurs when the kinematics result in the maximum amplification of the shed vortices in the wake, yielding maximum thrust per unit of input energy; this phenomenon has been termed ``wake resonance'' \cite{moored2012hydrodynamic}. The second theory \cite{wang2000vortex} argues that the preferred Strouhal number is connected with maximizing the angle of attack allowed, while avoiding the shedding of leading edge vortices.  The third \cite{saadat2017rules} holds that, for aquatic animals, the ratio of the tail beat amplitude to the body length essentially dictates the Strouhal number for cruise, since it requires a balance between thrust and drag. 

Here, we offer a simple alternative explanation for the observed peak in efficiency, and we also explain the rapid rise in efficiency at low $St$ and the more gradual decrease at high $St$.  Our explanation highlights the important role that fluid drag plays in determining the efficiency behavior. 

Consider a cruising animal, one that is moving at  constant velocity. We make the assumption that the thrust is produced primarily by its propulsor (for example, caudal fin for a fish, fluke for a mammal, wing for a bird), and that the drag is composed of two parts: the drag due to its body ($D_b$, proportional to the body surface area), and an ``offset'' drag due to its propulsor ($D_o$, proportional to the propulsor frontal area projected over its range of motion). More details are given below. 

This decomposition is illustrated in figure~\ref{fig:fishbirdfoil}, where the thrust-producing propulsor is separated from the drag-producing body and represented by an oscillating airfoil \cite{wu2011fish}. To be clear, fliers are distinct from swimmers in that fliers' propulsors need to produce lift to combat gravity, in addition to thrust to propel themselves forward. As far as steady forward cruising is concerned, however, the physics of forward propulsion is not affected by the additional requirement of lift \cite{wu2011fish}. 

We also simplify the motion of the propulsor to model it as a combination of heaving (amplitude $H$) and pitching (amplitude $\Theta$). Biologically-relevant motions are ones where the heaving and pitching motions are in phase or where the heaving motion leads the pitching motion by $90^\circ$ \cite{van2018scaling}. In cruise, our model requires that the thrust produced by the propulsor balances the total fluid drag experienced by the body and the propulsor. 

\begin{figure}
  \centering
  \includegraphics[width=0.5\linewidth]{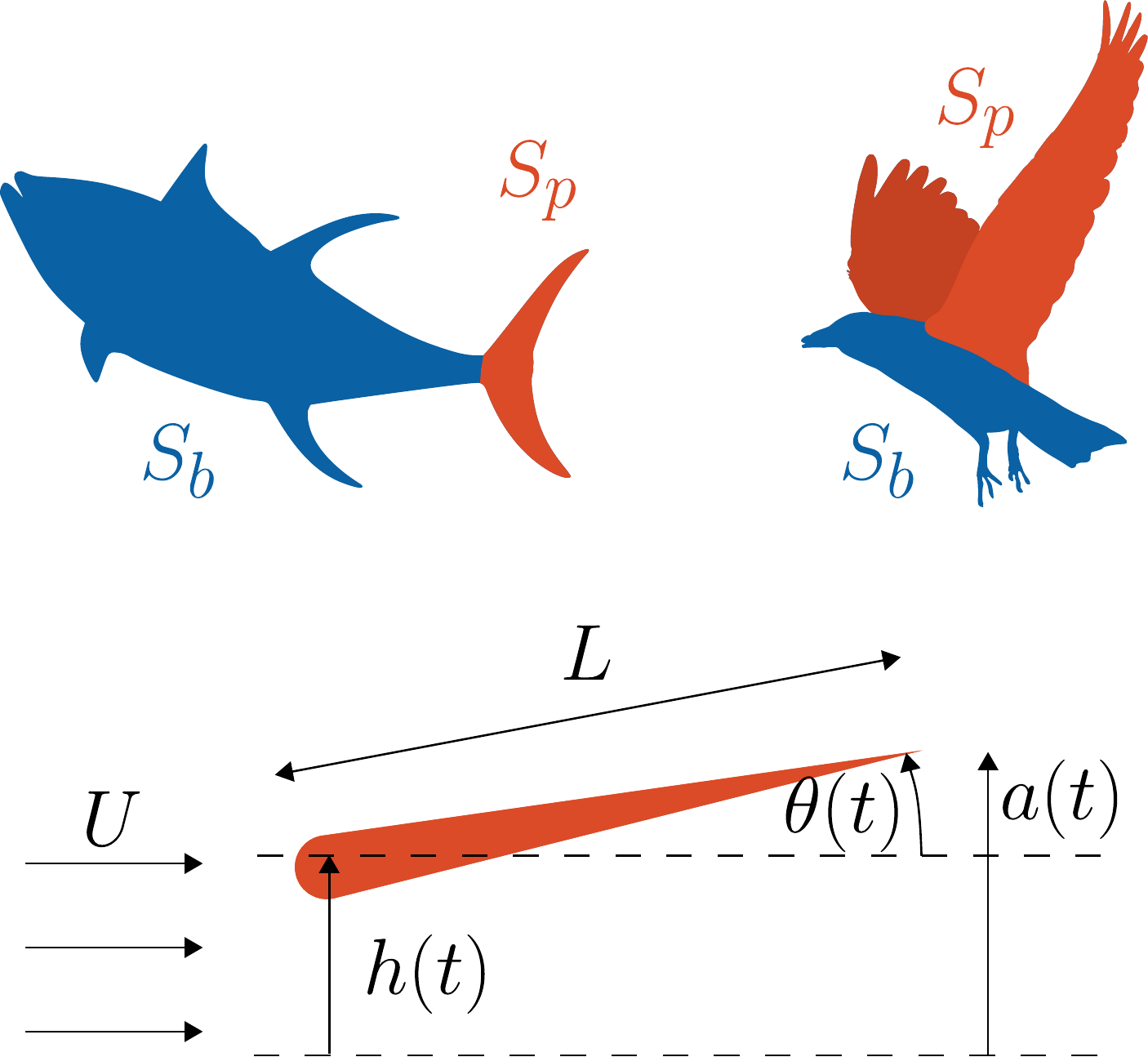}
  \caption{Swimmers and fliers can be decomposed into thrust-producing (orange) and drag-producing (blue) parts, with the propulsor aptly represented by an oscillating airfoil. }
  \label{fig:fishbirdfoil}
\end{figure}

We now consider the performance (thrust, power, and efficiency) of an isolated propulsor.  
For the net thrust $T$, we use the scaling 
\begin{equation}
  \label{eq1}
  T \sim \rho S_p V^2 -D_o,
\end{equation}
where $\rho$ is the density of the fluid, $S_p$ is the area of the propulsor, and $V$ ($\sim fA$) is the characteristic speed of the transverse motion of the propulsor. The $V^2$ scaling is derived in the SI Appendix, where it is also shown to be representative of biologically relevant flapping motions. In addition, the scaling is supported by theory \cite{garrick1936propulsion, lighthill1971large}, empirical curve fits on fish performance \cite{bainbridge1958speed, bainbridge1963caudal}, and the performance of a large group of swimming animals \cite{gazzola2014scaling}. As indicated above, we will assume that for a cruising animal the net thrust of the propulsor balances the drag of the body $D_b$, where $D_b \sim \rho S_b U^2$, and $S_b$ is the surface area of the body. Hence, for a negligible offset drag, 
\begin{equation}
  \label{eq:bal}
  St^2 \sim {S_b}/{S_p}.
\end{equation}
Previous work has proposed that this thrust-drag balance alone yields a constant Strouhal number \cite{gazzola2014scaling}.  However, \eqref{eq:bal} shows that this conclusion implicitly assumes that $D_o=0$ and that the area ratio $S_b/S_p$ remains constant, which will not hold across the many different species that cruise in the preferred range $0.2 < St < 0.4$.  To arrive at a more general result, we need to understand the energetics determining swimming and flying. The net thrust of the propulsor at peak efficiency then sets the cruising speed.

For the power expended, we adopt the scaling
\begin{equation}
  \label{eq2}
  P \sim \rho S_p f L \left( V^2 - V_h V_\theta \right),
\end{equation}
where $L$ is a characteristic length scale of the propulsor, and $V_h$ and $V_\theta$ are the transverse velocity scales characteristic of the heaving and pitching motions, respectively. This scaling is derived in the SI Appendix, where further details are given. It is based on established theory and analysis \cite{garrick1936propulsion, theodorsen1935general, sedov1965two}, and it is corroborated by a large set of experiments \cite{van2018scaling}. It derives from the nonlinear interaction of the power produced by the propulsor velocity and its acceleration, an interaction that is critical to our understanding of the large-amplitude motions observed in nature. 

We now consider the offset drag---that is, the drag of the propulsor in the limit of vanishing $f$---which scales as
\begin{equation}
  \label{eq3}
  D_o \sim \rho U^2 S_p g(\Theta).
\end{equation}
Here, $\Theta$ is the amplitude of the pitching motion, and the function $g(\Theta)$ is positive when $\Theta = 0$ and increases with $\Theta$ \cite{floryan2017scaling, van2018scaling}. 
The offset drag can be viewed as scaling with the projected frontal area of the propulsor, as in bluff body flows \cite{white2011fluid}. 

Hence, we arrive at 
\begin{equation}
  \label{eq4}
  \eta = \frac{T U}{P} \sim \frac{V^2 U - b_1 U^3 g}{ f L (V^2 - V_h V_\theta)},
\end{equation}
where the constant $b_1$ sets the relative importance of the drag term compared with the thrust term (in general, we expect $b_1$ to be a function of Reynolds number $Re=\rho L U/\mu$, where $\mu$ is the fluid viscosity). The efficiency can be recast in terms of the Strouhal number $St = 2fA/U$ and a dimensionless amplitude $A^* = A/L$, so that
\begin{equation}
  \label{eq5}
  \eta \sim \frac{ A^* \left( St^{2} - b_1g \right) }{St^{3} \left( 1 -  H^*  \Theta^* \right)}.
\end{equation}
The other nondimensional terms, $H^*=H/A$ and $\Theta^*=L\Theta/A$, represent, respectively, the amplitudes of the heaving and pitching motions relative to the total amplitude of motion.

We see immediately that to achieve high efficiency, the dimensionless amplitude $A^*$ should be large. This observation is consistent with the argument put forth by R. M. Alexander, where he proposed that large-amplitude motions are more efficient than small-amplitude motions \cite{alexander2003principles}. However, there are two potential limiting factors. First, as $A^*$ becomes larger, the instantaneous angle of attack increases, dynamic stall effects may become important, and the drag model given here for $D_o$ will be invalidated.
Second, animal morphology naturally sets a limit as to how large they can make $A^*$. For efficient cruising, therefore, $A^*$ should be as large as an animal's morphology allows, while avoiding dynamic stall at all times. Our argument is consistent with the experimental observations made by Saadat et al. \cite{saadat2017rules} in what we called the third theory.  The author of \cite{wang2000vortex} (the second theory) similarly argues for large-amplitude motions, although she argues that large-amplitude motions are connected to the optimal Strouhal number, whereas we argue that, all else fixed, the amplitude sets the total efficiency, but it does not dictate the optimal Strouhal number.

\begin{figure}
  \centering
  \includegraphics[width=0.5\linewidth]{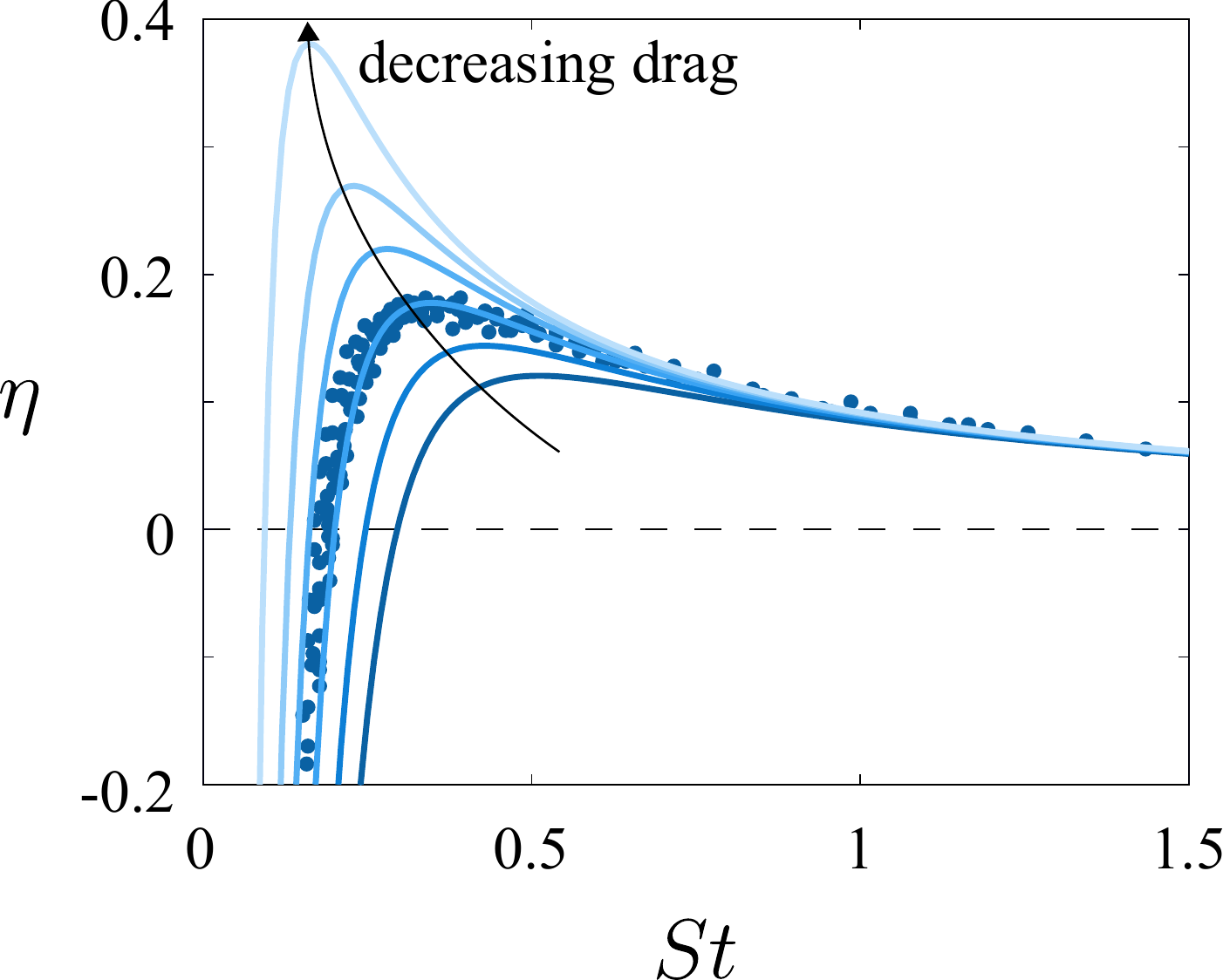}
  \caption{Efficiency $\eta$ as a function of $St$. Data are as given in figure~\ref{fig:eff} for a heaving and pitching NACA0012 foil \cite{quinn2015maximizing}. Solid lines are given by \eqref{eq5} with a fixed proportionality constant of 0.155. The drag constant, $b_1$, is set to 0.5, 0.35, 0.23, 0.15, 0.1, and 0.05 as the colors vary from dark to light, and we have set $g(\Theta) = \Theta$. The proportionality constant and the value of $b_1$ corresponding to the experimental data were calculated by a total least squares fit to the data. 
  }
  \label{fig:eff2}
\end{figure}

What about the optimal Strouhal number? When there is no offset drag ($b_1 = 0$), the efficiency increases monotonically as $St$ decreases, and the optimal efficiency is achieved in the limit $St \rightarrow 0$. However, in the presence of offset drag ($b_1 \neq 0$), the efficiency will become negative as $St \rightarrow 0$ because the drag dominates the thrust produced by the propulsor. In general, \eqref{eq5} gives negative  efficiencies at low $St$, a rapid increase with $St$ to achieve a positive peak value at $St = \sqrt{3b_1 g}$, and a subsequent slow decrease with further increase in $St$ as the influence of drag becomes weaker. The comparison between the form given by \eqref{eq5} and the data originally shown in figure~\ref{fig:eff} makes this clear, as displayed in figure~\ref{fig:eff2}.  The offset drag is crucial in determining the low $St$ behavior and in setting the particular $St$ at which the peak efficiency occurs. Note that the maximum value of the efficiency is directly related to the value of the drag constant $b_1$, which further emphasizes the critical role of the drag term in determining the efficiency behavior. The amplification of shed vortices described in the wake resonance theory (the first theory) may simply arise as a signature of the efficient production of net thrust, but this is purely speculative. 

Finally, we consider the composition of the motion---that is, the relative amounts of heaving and pitching. As shown in the SI Appendix, for biologically relevant flapping motions, the denominator of \eqref{eq5} is minimized (and hence efficiency is maximized) when $H = L\Theta$.  In other words, optimally efficient propulsors should have heaving and pitching motions that contribute equally to the total motion. When we also take the numerator of \eqref{eq5} into account, we actually expect the heaving contribution to be a little larger because the offset drag is dominated by pitch. We are not aware of biological measurements that would allow us to test the optimal heaving and pitching balance, so at this point it remains a hypothesis. 

We leave the reader with a final thought. We expect that the relative importance of the drag, captured by $b_1$, will depend on the Reynolds number.  Our drag model is similar to that for a bluff body, such as a sphere or cylinder, so we expect $b_1$ will be large at small Reynolds numbers and decrease as the Reynolds number increases until it reaches about 1000, above which the drag will be almost constant (at least for $Re<2 \times 10^5$, although biological measurements imply that the drag may remain constant up to $Re = 10^8$)\cite{schlichting1979boundary,gazzola2014scaling}. Therefore, at low Reynolds numbers, the location of the peak efficiency will change with Reynolds number: as Reynolds number increases, the optimal $St$ will decrease, until $b_1$ reaches its asymptotic value at a sufficiently high Reynolds number.  Our conclusion is consistent with biological measurements (at least for swimmers), where the preferred Strouhal number appears to decrease as the Reynolds number increases, until it reaches an asymptotic value \cite{gazzola2014scaling}. This further substantiates our claim that the presence of fluid drag on the propulsor is the crucial factor in creating an efficiency peak, which dictates the cruising conditions of swimming and flying animals. In other words, energetic considerations set the kinematics of the propulsor to the most efficient one, and the net thrust of the propulsor at peak efficiency  balances the drag of the body to set the cruising speed.

\subsection*{Materials and Methods}
The experimental setup is the same as described by Van Buren et al. \cite{van2018scaling}. Experiments on a heaving and pitching airfoil were conducted in a water tunnel with a $0.46 \times 0.3 \times 2.44$ m test section, with the tunnel velocity set to $U = 0.1$ m/s. A teardrop airfoil of chord $L = 0.08$ m, thickness 0.008 m, and span 0.279 m was used, yielding a chord-based Reynolds number of $Re = 8000$. 

Heaving motions were generated by a linear actuator (Linmot PS01-23 $\times$ 80F-HP-R), pitching motions about the leading edge were generated by a servo motor (Hitec HS-8370TH), and both were measured by encoders. The heaving and pitching motions were sinusoidal, as described in the SI Appendix, with frequencies $f = 0.2$ to 0.8 Hz every 0.1~Hz, heaving amplitudes $H = 0.01, 0.02, 0.03$ m, pitching amplitudes $\Theta = 5^\circ, 10^\circ, 15^\circ$, and phase angles $\phi = 0^\circ$ and $90^\circ$, with experiments performed on all combinations of the kinematic parameters. 

The forces and moments imparted by the water on the airfoil were measured by a six-component sensor (ATI Mini40) at a sampling rate of 100 Hz. The force and torque resolutions were $5 \times 10^{-3}$~N and $1.25 \times 10^{-4}$ N$\cdot$m, respectively, in the streamwise and cross-stream directions, and $10^{-2}$ N and $1.25 \times 10^{-4}$ N$\cdot$m, respectively, in the spanwise direction. Each case was run for 30 cycles, with the first and last five cycles used for warmup and cooldown. All sensors was zeroed before every case.

\subsection*{Acknowledgements}
This work was supported by Office of Naval Research Grant N00014-14-1-0533 (program manager Robert Brizzolara).

\bibliographystyle{unsrt}

\newpage

\appendix

\section{Supplementary Information}

\subsection*{Thrust and power}
Here we derive simple expressions for the mean thrust and power, as used in the main text, by considering sinusoidal heaving and pitching motions described by
\begin{align}
  h(t) &= H\sin(2\pi f t), \label{eq:heave}\\
  \theta(t) &= \Theta\sin(2\pi f t + \phi), \label{eq:pitch}
\end{align}
where pitch leads heave by a phase angle $\phi$. In our previous work \cite{van2018scaling}, we used aerodynamic theory to derive the following expressions for the mean thrust and power coefficients produced by a heaving and pitching foil:
\begin{align}
  C_T &= c_1 St^2 + c_2 St_h \Theta \sin \phi + c_3 St_\theta \Theta - c_4 \Theta, \label{eq:sm1}\\
  C_P &= c_5 St^2 + c_6 f^* St_h St_\theta \sin \phi + c_7 St_h \Theta \sin \phi  + c_8 f^* St_h^2 + c_9 f^* St_\theta^2 + c_{10} St_\theta \Theta, \label{eq:sm2}
\end{align}
where $St_h = 2fH/U$, $St_\theta = 2fL\Theta/U$, and the reduced frequency $f^* = fL/U$.  Also, $C_T= 2T/\rho S_p U^2$ is the thrust coefficient, and $C_P= 2P/\rho S_p U^3$ is the power coefficient.  Note that the term $c_4 \Theta$ represents the drag coefficient for the propulsor. These expressions were shown to collapse experimental data on a simple teardrop foil for all values of $\phi$.  

\begin{figure}
  \centering
  \includegraphics[width=\linewidth]{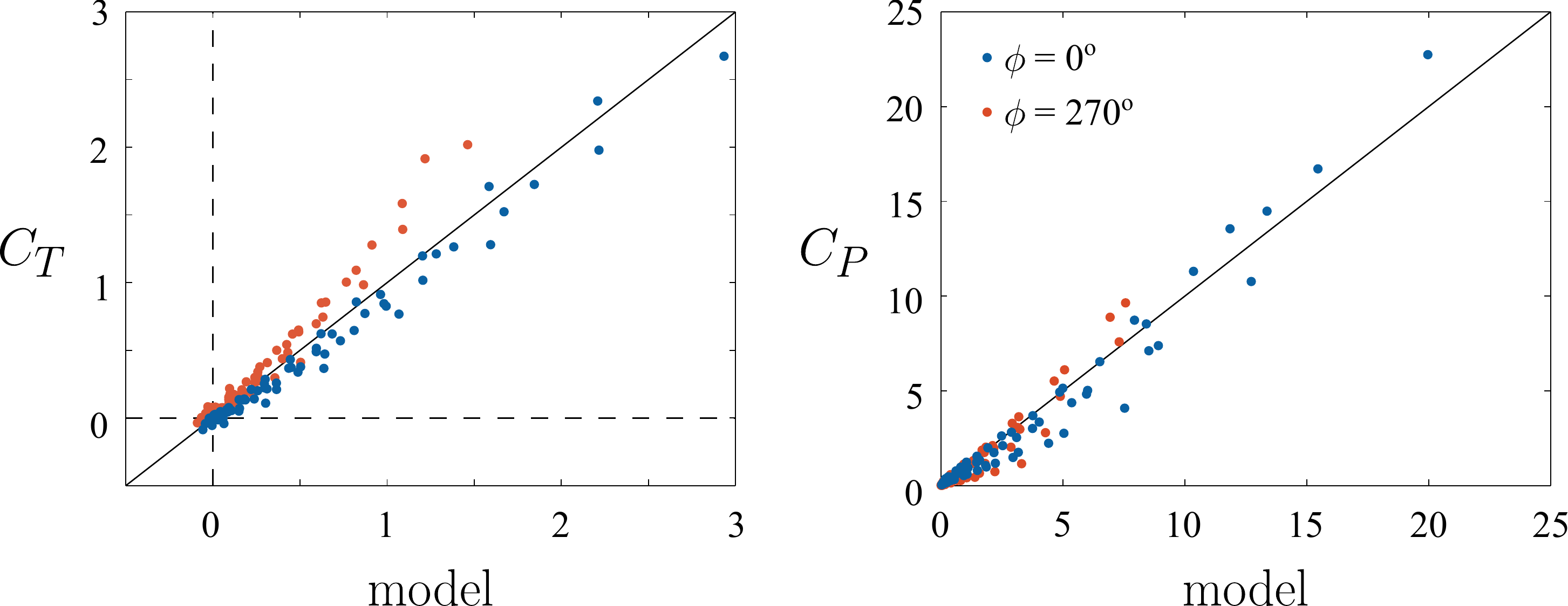}
  \caption{Thrust and power data plotted against expressions \eqref{eq:sm9}--\eqref{eq:sm10} for $\phi = 0^\circ$ (blue) and $\phi = 270^\circ$ (orange). The coefficients are $c_1 = 4.65$, $c_4 = 0.49$, $a_2 = 62.51$.}
  \label{fig:thrustpower_reduced3}
\end{figure}

For the biologically-relevant phase angles $\phi = \{0^\circ, 270^\circ\}$, we find that the $c_2$ and $c_3$ terms in thrust, and the $c_{10}$ and $c_7$ terms in power, are small relative to the other terms and can be neglected.  
For power, we use $St^2 = St_h^2 + St_\theta^2 + 2 St_h St_\theta \cos \phi$.  As a result, we now propose, for $\phi = \{0^\circ, 270^\circ\}$,
\begin{align}
C_T &= c_1 St^2 - c_4\Theta, \label{eq:sm7} \\
  C_P &= a_1 St^2 + a_2 f^* St^2 + a_3 f^* St_h St_\theta. \label{eq:sm8}
\end{align}
We have introduced new constants $a_i$ to avoid confusion with the previous constants $c_i$ in the power. All signs have been absorbed into the constants. Note that we now have the same thrust and power expressions for both phases. 
Based on the numerical values of the constants in \eqref{eq:sm7}--\eqref{eq:sm8}, as found from the experimental data, we can propose a further reduction, where
\begin{align}
 C_T &= c_1 St^2 - c_4\Theta, \label{eq:sm9} \\
  C_P &= a_2 f^* \left( St^2 - St_h St_\theta \right). \label{eq:sm10}
\end{align}
Plotting the thrust and power data against expressions \eqref{eq:sm9}--\eqref{eq:sm10} yields figure~\ref{fig:thrustpower_reduced3}. The collapse using these reduced models is as good as obtained by Van Buren et al. \cite{van2018scaling} using the full expressions given by \eqref{eq:sm1} and \eqref{eq:sm2}. 

Equations~\eqref{eq:sm9} and \eqref{eq:sm10} can be written dimensionally, so that, for $\phi = \{0^\circ, 270^\circ\}$,
\begin{align}
T &\sim \rho S_p V^2 - D_o, \label{eq:sm11} \\
  P &\sim \rho S_p f L(V^2 - V_h V_\theta), \label{eq:sm12}
\end{align}
where $D_o$ is the drag offset. 

\subsection*{Motion composition}
The total amplitude for a motion with arbitrary phase is
\begin{equation}
  \label{eq:c1}
  A^2 = H^2 + 2HL\Theta\cos\phi + L^2 \Theta^2.
\end{equation}
For biologically-relevant phases, we then have
\begin{align}
  \phi = 0^\circ: A &= H + L\Theta,\\
  \phi = 270^\circ: A^2 &= H^2 + L^2\Theta^2.
\end{align}
For both phases, $HL\Theta/A^2$ is maximized when $H = L\Theta$, minimized when one of them is zero, and always less than 1. (This can be shown using calculus for $\phi = 0^\circ$, and using right triangles for $\phi = 270^\circ$.) So for both phases, the denominator of \eqref{eq5} is positive, and is minimized when the heave and pitch amplitudes are equal.

\end{document}